\documentclass[%
reprint,
superscriptaddress,
showpacs,
 amsmath,amssymb,
pre,
floatfix,
]{revtex4-1}

\usepackage{graphicx}
\usepackage{dcolumn}
\usepackage{bm}
\usepackage{hyperref}

\newcommand{\sgn}[1]{\mbox{sgn}#1}
\newcommand{\PR}{\textit{Phys. Rev. }}
\newcommand{\PRL}{\textit{Phys. Rev. Lett.}}
\newcommand{\JMP}{\textit{J. Math. Phys.}}
\newcommand{\JSTAT}{\textit{J. Stat. Mech. (Theor. Exp.)}}
\newcommand{\rmd}{\text{d}}

\begin{document}


\title{Spin-oscillator model for DNA/RNA unzipping by mechanical force} 

\author{A.~Prados}
\affiliation{%
 F\'{\i}sica Te\'orica, Universidad de Sevilla\\
 Apartado de Correos 1065, E-41080 Sevilla, Spain
}%


\author{A.~Carpio}
\affiliation{%
Departamento de Matem\'atica Aplicada, Universidad Complutense de Madrid, 28040 Madrid, Spain
}
\author{L.~L.~Bonilla}
\affiliation{%
G.~Mill\'an Institute for Fluid Dynamics, Nanoscience and Industrial Mathematics, Universidad Carlos III de Madrid, 28911 Legan\'es, Spain
}%


\date{\today}

\begin{abstract}
We model unzipping of DNA/RNA molecules subject to an external force by a spin-oscillator system. The system comprises a macroscopic degree of freedom, represented by a one-dimensional oscillator, and internal degrees of freedom, represented by Glauber spins with nearest-neighbor interaction and a coupling constant proportional to the oscillator position. At a critical value $F_c$ of an applied external force $F$, the oscillator rest position (order parameter) changes abruptly and the system undergoes a first-order phase transition. When the external force is cycled at different rates, the extension given by the oscillator position exhibits a hysteresis cycle at high loading rates whereas it moves reversibly over the equilibrium force-extension curve at very low loading rates. Under constant force, the logarithm of the residence time at the stable and metastable oscillator rest position is proportional to $(F-F_c)$ as in an Arrhenius law.
\end{abstract}

\maketitle


\section{Introduction}

Many physical situations can be modelled by a mechanical system coupled to a thermal bath or to spin systems. Examples abound, the collective Jahn-Teller effect has been analyzed by spin-phonon systems \cite{fed73,rik77,rik78}, mass spectrometry through a nanoelectromechanical oscillator whose resonant frequency decreases as single molecules are added thereto \cite{boi09}, decoherence of a spin representing a two-level system due to coupling to a boson bath (the spin-boson system) \cite{leg87}, a classical oscillator coupled to a spin causes wave function collapse thereof \cite{bon92}, a 1/2-spin representing a nonlinear Josephson phase quantum bit is coupled to an oscillator (superconducting resonator) and to  a classical signal \cite{hof09,oco10}, rippling in clamped graphene sheets has been investigated by means of a spin-string system \cite{BCPyR12}, etc.

Very recently, we have introduced a simple model in which a single oscillator is coupled to a chain of Ising spins undergoing Glauber dynamics in contact with a thermal bath \cite{PByC10,BPyC10,PByC11a}. In our model, the spins in the chain are coupled only to their nearest neighbors, but their coupling constant is proportional to the oscillator position, which makes their interaction effectively long ranged. In equilibrium, elimination of the oscillator coordinates gives rise to an effective spin interaction equivalent to a one dimensional Ising model with mean field coupling \cite{BW34}. There is a second order phase transition at a finite temperature, with the oscillator rest position as its order parameter. Above the critical temperature, the oscillator rest position is zero, thereby coinciding with that of the uncoupled oscillator. Below the critical temperature, two symmetric nonzero rest positions issue forth symmetrically from zero as in the case of a pitchfork bifurcation. In the limit of fast relaxation of the spins compared to the natural period of the oscillator, the oscillator position satisfies an effective equation having  both nonlinear force and  nonlinear friction terms \cite{PByC10,PByC11a}. Interestingly, this nonlinear friction arises from the coupling of the macroscopic elastic mode with the internal degrees of freedom (modelled in our system by the spins). A related mechanism has been proposed to explain the ``internal friction'' observed in experiments with proteins or polymers in solution \cite{ZyZ02}.

In recent years, technological development has allowed to manipulate or visualize individual molecules and to measure microscopic forces with high precision instruments. These single-molecule experiments (SME) provide key information about the thermodynamic and kinetic properties of biomolecules, offering a complementary but different perspective to understand molecular processes.  An extensive review of these techniques can be found in \cite{Ri06}. Using SME, distributions describing certain
molecular properties can be measured, thereby allowing to characterize the kinetics of biomolecular reactions and the observation of  possible intermediate states. A typical outcome of SME are the force-extension curves for DNA, RNA and other biomolecules. In a seminal paper, Liphardt et al.~\cite{LOSTyB01} pull the nucleic acid molecule by an increasing applied force until it unfolds at a critical value of the force, $F_c\simeq 14.5$pN. Afterwards, the molecule is pushed back, by decreasing the force, until it refolds. At low pulling (pushing) rates, the stretching and relaxing force-extension curves are superimposed, and
the molecule unfolds at the critical value of the force $F_c$. Taking a closer look at this transition, hopping between the two possible extension values is observed. This suggests that the system is bistable: there are two possible states of the molecule with stochastic transitions between them. This physical picture is confirmed by experiments carried out at constant load, in a narrow region around the critical force (see fig.\ 2(C) of ref. \cite{LOSTyB01}). When cycles of pulling/pushing the molecule are carried out at high loading rates, the extension of the molecule occurs at a higher force $F_+>F_c$, whereas the hairpin folds at a lower force $F_-<F_c$. Thus a hysteresis cycle arises, and some authors have been claimed this to be a signature of irreversible non-equilibrium behavior \cite{LOSTyB01,LDSTyB02,Ri03,CRJSTyB05}.  More recently many works have tried to understand these unzipping  experiments from a physical point of view \cite{LDSTyB02,Ri03,CRJSTyB05,WyR02,MyR05,WMLSBRyT07,MWLSBTyR07,HFyR09,Hu10,GLOBPyW11}.

In this paper, we add an external force to our previous oscillator-spin model \cite{PByC10} and analyze the resulting force-extension curves. Qualitatively, these curves have the same features as those of the force-extension curves measured in experiments with DNA, RNA and other biopolymers described above. At subcritical temperatures, our spin-oscillator system has a first order phase transition at a critical force $F_c$ with the oscillator rest state as its order parameter. We find that the DNA force-extension curves correspond to cycling at different rates the curves of the first order phase transition. As in the experiments, we find a region of metastability in a certain range of forces, close enough to $F_c$. Moreover, the residence time spent at the basin of attraction of both the stable and metastable states obey an Arrhenius law: its logarithm is proportional to $F-F_c$.

The rest of the paper is as follows. In Section \ref{s2}, the oscillator-spin model is motivated in a biological context and its equilibrium properties analyzed. The dynamical behavior of the model is analyzed in Section \ref{s3}. The oscillator obeys Newton's second law with a mean-field force due to the coupling with the spins. The latter flip stochastically following Glauber dynamics at temperature $T$ \cite{Gl63}. This causes the oscillator position to become a stochastic process. In the limit of fast spins, the spin-oscillator coupling gives rise to a nonlinear friction term, which drives the oscillator to equilibrium. In Section \ref{s4}, we present Monte Carlo simulations of the system and analyze them in the light of the effective potential acting on the oscillator. Section \ref{s5} contains the main conclusions of the present work.

\section{\label{s2}The model}

\begin{figure}[b]
\centerline{\includegraphics[width=\linewidth]{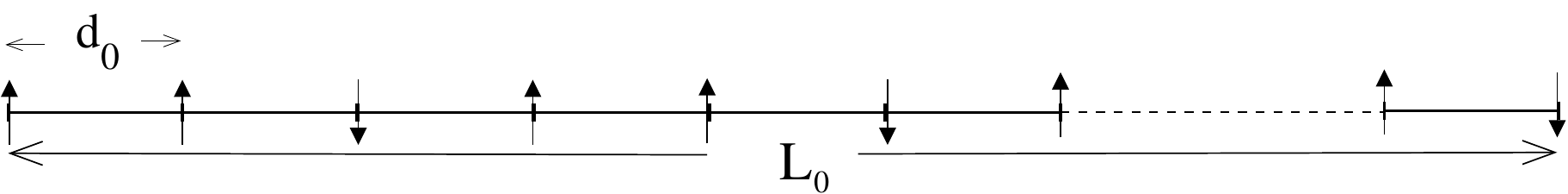}}
\caption{\label{fig3} Sketch of the model described in the main text. The spin representing the internal degree of freedom at each lattice site is shown.}
\end{figure}
We consider a one dimensional chain of length $L_0$. There is a large number $N+1$ of ``internal'' degrees of freedom sitting at regularly spaced lattice sites that are modeled by Ising spins. Thus, the distance between spins is $d_0=L_0/N$ (see fig. \ref{fig3}). Assume that we stretch the chain so that its length becomes $L=L_0+\Delta$. For the sake of simplicity, we will assume that the spins are regularly spaced after the stretching, so that the distance between two neighboring spins changes to $d=d_0+\Delta/N$. This assumption amounts to a ``mean field'' approximation. The potential energy of the system is
\begin{subequations}
\begin{equation}\label{1.1a}
    {\cal V}(\Delta,\bm{\sigma}) =\frac{1}{2}m \omega^2 \Delta^2+J(\Delta)\sum_{i=1}^{N+1}\sigma_i\sigma_{i+1},
\end{equation}
 where
\begin{equation}\label{1.1b}
     J(\Delta)=J_0-\mu \Delta ,
\end{equation}
\end{subequations}
is a function of $\Delta$. The potential ${\cal V}$ contains a harmonic macroscopic elastic term $m\omega\Delta^2/2$ and a spin energy arising from a nearest-neighbor interaction. The spin coupling constant $J$ depends linearly on the separation between sites, and equals $J_0$ for the initial chain length $L_0$. This simple choice is reasonable for $\Delta\ll L$. The interaction between nearest neighbor spins mimics (in a very simple way) the short-ranged interaction between the internal degrees of freedom of complex biological molecules like nucleic acids. We assume that both $J_0$ and $\mu$ are positive. Let us define
\begin{equation}\label{1.2}
    x=\Delta-\frac{J_0}{\mu},
\end{equation}
such that $J(x)$ vanishes for $x=0$. For $x<0$ (folded state) the interaction between the spins is ferromagnetic, while for $x>0$ (unfolded state) it is antiferromagnetic. In terms of $x$, ${\cal V}$ becomes
\begin{equation}\label{1.3}
    {\cal V}=\frac{1}{2}m\omega^2 x^2+F_c x-\mu x \sum_{i=1}^{N+1} \sigma_i \sigma_{i+1}, \quad F_c=\frac{m \omega^2 J_0}{\mu}.
\end{equation}
except for an irrelevant additive constant. The parameter $F_c$ has the dimensions of a force. If an external load $F$ is applied to the system, a new term $-Fx$ is added to (\ref{1.3}) so that the potential energy is now
\begin{equation}\label{1.4}
    {\cal V}=\frac{1}{2}m\omega^2 x^2 - H x -\mu x \sum_{i=1}^{N+1} \sigma_i \sigma_{i+1}, \quad H=F-F_c.
\end{equation}
This potential energy is the same as introduced in ref.\ \cite{PByC10,BPyC10,PByC11a}, except for the extra term $-Hx$. Interestingly, in the ``zero field'' case, our system has a second order phase transition at a critical temperature $T_c$, given by \cite{PByC10,BPyC10,PByC11a}
\begin{equation}\label{1.5}
    T_c=\frac{\mu^2 (N+1)}{m\omega^2 k_B}
\end{equation}
where $k_B$ is Boltzmann's constant. The rest position $x$ is the order parameter of the transition: for $T>T_c$, $x=0$, whereas for $T<T_c$ there are two equally probable equilibrium states with rest positions $x=\pm x_0$ ($x_0>0$).

The potential energy (\ref{1.4}) has the key ingredients to model DNA/RNA behavior in unfolding/refolding experiments. For $T<T_c$ and applied force  $F<F_c$, $H<0$ stabilizes the solution with rest state $-x_0$ (folded state), whereas for$F>F_c$, $H>0$ and the stable solution has rest state $+x_0$ (unfolded state). The main effect of the ``external field'' $H$ is that the system undergoes a first order phase transition at $H=0$ ($F=F_c$) \cite{Landau}. Thus, $F_c$ is the a critical value of the force $F$: at any temperature $T<T_c$ the oscillator rest position as a function of the force changes abruptly at $F=F_c$. Furthermore there is a region of metastability around $F_c$, as discussed below in Section \ref{s2b}.

To analyze our model, it is convenient to render its equations of motion dimensionless first. Let $\omega^{-1}$ be the time unit. The elastic and the spin term in the potential are of the same order if the scale of the oscillator position is $[x]=\mu (N+1)/(m\omega^2)$. The force and the spin term are of the same order provided the scale of $F$ is $[F]=\mu (N+1)$, and therefore the order of magnitude of the potential is $[F][x]=\mu^2 (N+1)^2/(m\omega^2)=(N+1)k_B T_c$. This scaling is reasonable because makes $\mathcal{V}$ extensive. For the critical temperature $T_c$ to be size-independent, we must assume that $\mu$ scales as $(N+1)^{-1/2}$ in the limit of large system size \cite{PByC10,BPyC10,PByC11a}.
\begin{table}[ht]
\begin{center}\begin{tabular}{ccccccc}
 \hline
 $x$ &  $t$ &  $F$   & ${\cal V}$ & $\theta$  \\
$\frac{\mu (N+1)}{m\omega^2}$ & $\frac{1}{\omega}$ & $\mu(N+1)$ & $(N+1) k_B T_c$
& $\frac{T}{T_c}$ \\
 \hline
\end{tabular}
\end{center}
\caption{Nondimensional units and parameters.}
\label{table1}
\end{table}
Thus we can define nondimensional variables according to $x^*=x/[x]$,
$t^*=t/[t]$, $\mathcal{V}^*=\mathcal{V}/[\mathcal{V}],\ldots$, where the units $[x]$, $[t]$, $[\mathcal{V}],\ldots$ are as defined in Table \ref{table1}. The dimensionless potential is
\begin{subequations}\label{1.6}
\begin{equation}\label{1.6a}
   {\cal  V}^*=\frac{\cal V}{(N+1)k_B T_c}=\frac{x^{*2}}{2}-H^* x^*-\frac{x^*}{N+1} \sum_{i=1}^{N+1} \sigma_i \sigma_{i+1},
\end{equation}
\begin{equation}\label{1.6b}
   H^*=F^*-F_c^*,
\end{equation}
\end{subequations}
with
\begin{equation}\label{1.7}
    F_c^*=\frac{F_c}{\mu (N+1)}=\frac{m\omega^2 J_0}{\mu^2 (N+1)}=\frac{J_0}{k_B T_c}.
\end{equation}
We will drop the asterisks in the following (so as not to clutter our formulas), and from now on every expression will be written in terms of the dimensionless variables and parameters.

\subsection{\label{s2b}Equilibrium state. Effective potential}

In equilibrium, the joint probability distribution for the oscillator position $x$ and the spin configuration $\bm{\sigma}=\left\{ \sigma_1,\ldots,\sigma_{N+1} \right\}$ is the canonical distribution which, in nondimensional variables, is
\begin{equation}\label{2.1}
    {\cal P}_{\text{eq}}(x,\bm{\sigma})=\frac{1}{Z} \exp \left[ - (N+1) \mathcal{V}(x,\sigma)/\theta \right] .
\end{equation}
Here $Z$ is a normalization constant. Let us study the equilibrium values of the oscillator position $x$. Then, we sum over the spin variables to obtain the marginal distribution probability
\begin{equation}\label{2.2}
    {\cal P}_{\text{eq}}(x)=\sum_{\bm{\sigma}}  {\cal P}_{\text{eq}}(x,\bm{\sigma})=\frac{1}{\tilde{Z}}
     \exp \left[ - (N+1) \mathcal{V}_{\text{eff}}(x)/\theta \right],
\end{equation}
where $\tilde{Z}=2^N Z$. In Eq. (\ref{2.2}), $\mathcal{V}_{\text{eff}}$ is an effective potential for the $x$ variable,
\begin{equation}\label{2.3}
    \mathcal{V}_{\text{eff}}(x)=\frac{x^2}{2}-H x- \theta \,\ln\cosh\left(\frac{x}{\theta}\right),
\end{equation}
whose minima will be the stable equilibrium values of $x$. Therefore,
\begin{equation}\label{2.4}
    x_{\text{eq}}=H+\tanh\left(\frac{x_{\text{eq}}}{\theta}\right),
\end{equation}
gives the oscillator rest position $x_{\text{eq}}$ in equilibrium as a function of the dimensionless external field $H=F-F_c$ and temperature $\theta$.
\begin{figure}
\includegraphics[width=\linewidth]{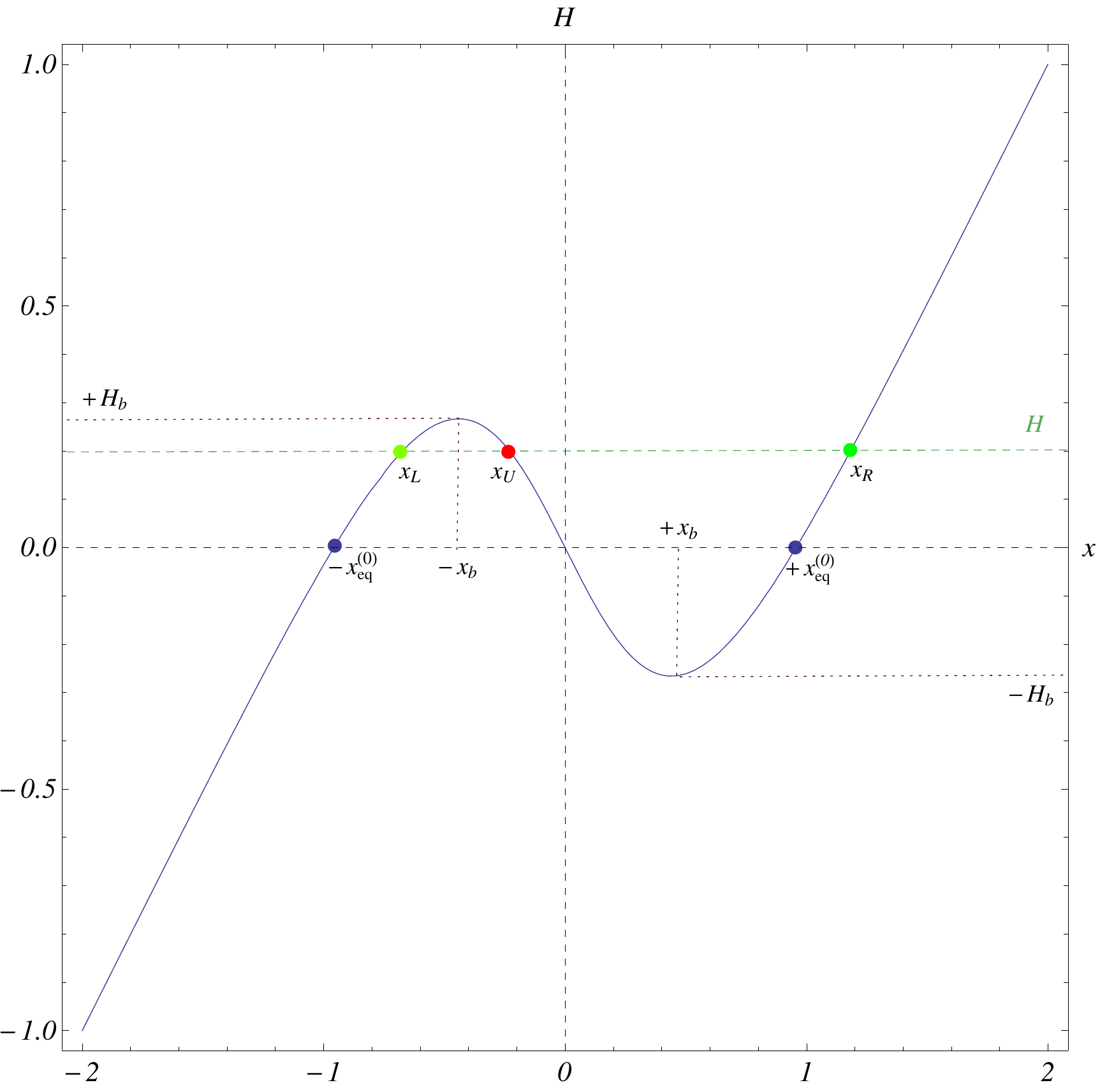}
\caption{\label{fig4} (Color online) Plot of $H$ vs $x_{\text{eq}}$ for $\theta=0.5$. The values $\pm H_b$ between which there are three equilibrium values of the oscillator position are shown. For a given value of $H$, $-H_b<H<+H_b$, the two locally stable equilibrium points $x_L$ and $x_R$ (green) and the unstable one $x_U$ (red) are indicated.  The two symmetric stable equilibrium points $\pm x_{\text{eq}}^{(0)}$ corresponding to the zero field case are also shown.  The qualitative shape of the curve is the same for all the subcritical temperatures $\theta<1$.}
\end{figure}

For $H=0$, we recover the model analyzed in refs. \cite{PByC10,BPyC10,PByC11a}, in which ${x}_{\text{eq}}=0$ is always a solution for any $\theta$. For $\theta>1$, it is the only solution, it corresponds to a maximum of ${\cal P}_{ \text{eq}}$ and is therefore stable. At $\theta=1$ two new stable equilibria corresponding to two different maxima of ${\cal P}_{ \text{eq}}$ bifurcate from that having ${x}_{\text{eq}}=0$. For $\theta<1$, the positions of these maxima are $\pm x_{\text{eq}}^{(0)}$. As $\theta\rightarrow 1^-$, we have
\begin{equation}\label{2.5}
    x_{\text{eq}}^{(0)} \sim \sqrt{3 \left(1-\theta\right)} \, .
\end{equation}
On the other hand, for $\theta\rightarrow 0^+$, we have $x_{\text{eq}}^{(0)}\rightarrow  1$, which is the maximum value of $x_{\text{eq}}^{(0)}>0$. For $H=0$, the two (positive or negative) equilibrium rest positions $\pm x_{\text{eq}}^{(0)}$ are equiprobable because they correspond to equally deep minima of the effective potential (\ref{2.3}), which is an even function of $x$ for $H=0$.

For $H\neq 0$ and temperatures below critical, $\theta<1$, the field term $-Hx$ in eq. (\ref{2.3}) breaks the symmetry between $x>0$ and $x<0$, the effective potential is no longer an even function of $x$. If $H<0$, the field term favors the negative branch $x<0$, since it gives a negative contribution to the effective potential. Therefore, we expect to find the system in the ``folded'' state $x<0$ for low values of the loading force, $F\ll F_c$. On the contrary, for $H>0$, the energy of the positive branch $x>0$ will be lowered by the field term, stabilizing it. Thus, the system will be in the ``unfolded'' state $x>0$ for high values of the loading force, $F\gg F_c$.

Differentiating Eq. (\ref{2.4}) with respect to $H$, we find
\begin{equation}\label{2.6}
    \partial_H x_{\text{eq}}=\left[ 1-\frac{1}{\theta}\, \mbox{sech}^2\! \left(\frac{x_{\text{eq}}}{\theta}\right)\!\right]^{-1}.
\end{equation}
The right hand side of this expression is the reciprocal of the second derivative of the effective potential, and therefore we can rewrite (\ref{2.6}) as
\begin{equation}\label{2.6a}
    \partial_{x}^2{\cal V}|_{x=x_{\text{eq}}} \partial_H x_{\text{eq}}=1 .
\end{equation}
Then the effective potential has a local minimum and the corresponding equilibrium rest position is stable for $\partial_H x_{\text{eq}}>0$, while the equilibrium rest state is unstable for $\partial_H x_{\text{eq}}<0$. Since $\partial_H x_{\text{eq}}=\theta/(\theta-1)$ for $x_{\text{eq}}=0$, the zero rest position of the oscillator is stable at $\theta>1$ and unstable at subcritical temperatures $\theta<1$.

The equilibrium position $x_{\text{eq}}$ is a monotonically increasing function of the applied load $F$ (or the applied field $H=F-F_c$) for supercritical temperatures, $\theta>1$. There is a unique value of $x_{\text{eq}}$ for each value of the applied force $F$ if $\theta>1$, and it is stable. As shown in Fig. \ref{fig4} and given by (\ref{2.6}), $x_{\text{eq}}$ is not monotonic for subcritical temperatures $\theta<1$: It has a local maximum at $-x_b$ and a local minimum at $+x_b$, with
\begin{equation}\label{2.7}
    \cosh^2 \left(\frac{x_b}{\theta}\right)=\frac{1}{\theta}.
\end{equation}
For each value of $H$ between $-H_b$ and $+H_b$, with
\begin{equation}\label{2.8}
    H_b=x_b-\tanh\left( \frac{x_b}{\theta} \right),
\end{equation}
there are three possible values of $x_{\text{eq}}$: $x_U$ between $-x_b$ and $x_b$ is therefore unstable, while the other two, $x_L<0$ and $x_R>0$ are locally stable. The absolute minimum of the potential corresponds to the value of $x$ having the same sign as the applied field $H$. The other local minimum, with $\sgn (x)\neq \sgn(H)$, is a metastable state in the thermodynamic sense. Then, we expect to find bi-stability in the system for $|H|<H_b$, i.e. for a given range of loadings $|F-F_c|<H_b$ around the ``critical'' force $F_c$.

\section{\label{s3}Dynamics}

The Hamilton equations of motion corresponding to the nondimensional Hamiltonian function
\begin{equation}\label{3.1}
    {\cal H}(x,p,\bm{\sigma})=\frac{p^2}{2}+{\cal V}(x,\bm{\sigma}),
\end{equation}
with potential energy given by Eq. (\ref{1.6a}), are
\begin{equation}\label{3.2a}
    \dot{x}=p , \quad \dot{p}=-\partial_x {\cal V}(x,\bm{\sigma}),
\end{equation}
so that
\begin{equation}\label{3.2b}
    \ddot{x}=-x+H+\frac{1}{N+1}\sum_{i=1}^{N+1}\sigma_i \sigma_{i+1}.
\end{equation}
According to the stochastic Glauber dynamics, the spins flip at a rate
\cite{Gl63}
\begin{subequations}
\begin{equation}\label{3.3a}
    W_i(x,\bm{\sigma})=\frac{\alpha}{2}\left[ 1-\frac{\gamma(x)}{2}\sigma_i \left(\sigma_i + \sigma_{i+1}\right)\right],
\end{equation}
\begin{equation}\label{3.3b}
    \gamma(x)=\tanh \left( \frac{2 x}{\theta} \right).
\end{equation}
\end{subequations}
Here, $W_i(x,\bm{\sigma})$ is the transition rate from configuration $\bm{\sigma}$ to $R_i\bm{\sigma}$, the same as $\bm{\sigma}$ except for the sign of the $i$-th spin. Since the oscillator evolution equation (\ref{3.2b}) includes a term that depends on the stochastically changing spin configuration, the oscillator position becomes an stochastic process.

The average values of the spin correlations,
\begin{equation}\label{3.4}
    C_{i,n}=\sigma_i \sigma_{i+n},
\end{equation}
satisfy the system of equations
\begin{eqnarray}
    \nonumber \frac{\rmd \langle C_{i,n}\rangle}{\rmd t}&=&\alpha \Big[-2\langle C_{i,n}\rangle+\frac{1}{2} \langle \gamma(x) \big( C_{i,n-1}+C_{i,n+1}  \\
    && \quad\quad + C_{i-1,n+1}+C_{i+1,n-1} \big) \rangle\Big], \label{3.5}
\end{eqnarray}
for $n\geq 1$, with the boundary condition $C_{i,0}=1$. If the oscillator position $x$ were time-independent, the spins would reach the equilibrium distribution corresponding to the constant value $x$. Then the average spin correlations $\langle C_{i,n}\rangle=\left[\tanh(x/\theta)\right]^n$ would be independent of $i$ due to the spatial translation invariance. Something similar occurs in the limit of large system size, $N\gg 1$. Both $x$ and the spin correlations $C_{i,n}$ become macroscopic self-averaging variables, i.e. they tend to their respective ``macroscopic values'', $\widetilde{x}$ and $\widetilde{C}_n$ (independent of $i$), which coincide with their averages and are the most probable values of the corresponding stochastic variables \cite{vk92}. Splitting both $x$ and the correlations $C_{i,n}$ in their corresponding macroscopic and fluctuating parts,
\begin{equation}\label{3.6}
  x = \widetilde{x}+ \delta x , \quad
  C_{i,n} = \widetilde{C}_n+ \delta C_{i,n}  ,
\end{equation}
where $\delta x$ and $\delta C_{i,n}$  are $\mathcal{O}(N^{-1/2})$, Eqs. (\ref{3.2b}), (\ref{3.5}) and (\ref{3.6}) yield the evolution equations,
\begin{subequations}\label{3.7}
\begin{equation}\label{3.7a}
    \ddot{\widetilde{x}} = -\widetilde{x}+H+\widetilde{C}_1,
\end{equation}
\begin{equation}\label{3.7b}
    \frac{\rmd \widetilde{C}_n}{\rmd t} = \alpha \left[-2\widetilde{C}_n+\gamma(\widetilde{x})
  \left(\widetilde{C}_{n-1}+\widetilde{C}_{n+1}\right)\right], \quad n \geq 1,
\end{equation}
\end{subequations}
to be solved with $\widetilde{C_0}=1$ and appropriate initial conditions. In Eq. (\ref{3.7b}) we have neglected terms or order $1/N$ such as $\langle (\delta x)^2\rangle$, $\langle \delta x \delta C_{i,n}\rangle$, etc.

As the spins represent the internal degrees of freedom of the molecule, we assume that they evolve rapidly compared with the time scale of the macroscopic degree of freedom modeled by the oscillator position. Thus, the dimensionless characteristic attempt rate satisfies $\alpha\gg 1$ (recall that the unit of time has been chosen as  $\omega^{-1}$). Thus, we can solve approximately the system of equations (\ref{3.7b}) using a power series in the small parameter $\alpha^{-1}$ \cite{ByP93,BPyR94,BPyS97}, a procedure akin to the Hilbert method in kinetic theory.  If $\widetilde{x}$ were time-independent, the spin correlations $\widetilde{C}_n$ would reach the equilibrium values corresponding to $\widetilde{x}$,
\begin{equation}\label{3.7c}
    \widetilde{C}_{n,\text{eq}}=\left[ \widetilde{C}_{1,\text{eq}} \right]^n, \qquad \widetilde{C}_{1,\text{eq}}=\tanh \left(\frac{\widetilde{x}}{\theta}\right)\!,
\end{equation}
in the long time limit. The leading order correction of this result is \cite{PByC10}
\begin{equation}\label{3.7d}
    \widetilde{C}_1=\widetilde{C}_{1,\text{eq}}-\tau \frac{\rmd \widetilde{C}_{1,\text{eq}}}{dt},
\end{equation}
where $\tau$ is the spins average relaxation time \cite{ByP93,BPyR94}
\begin{equation}\label{3.7e}
    \tau=\frac{1}{2\alpha} \frac{1+\widetilde{C_1}_{\text{eq}}^2}{\left(1-\widetilde{C_1}_{\text{eq}}^2 \right)^2}.
\end{equation}
Equation (\ref{3.7d}) does not depend on the initial condition for $\widetilde{C_1}$ which is forgotten after a time much shorter than the oscillator natural period. Inserting Eqs. (\ref{3.7c})-(\ref{3.7e}) into (\ref{3.7a}), we obtain
\begin{equation}\label{3.9}
\frac{\rmd^2\widetilde{x}}{\rmd t^{2}}+ \frac{1}{2\alpha\theta} \frac{1+\tanh^2(
\frac{\widetilde{x}}{\theta})}{1-\tanh^2(\frac{\widetilde{x}}{\theta})}\, \frac{\rmd
\widetilde{x}}{\rmd t}+\widetilde{x}-H- \tanh\left(\frac{\widetilde{x}}{\theta}\right)=0,
\end{equation}
which can be rewritten in terms of the nondimensional
effective potential (\ref{2.3}) and the friction coefficient
\begin{equation}\label{3.10}
R(\widetilde{x}) =\frac{1+\tanh^2( \frac{\widetilde{x}}{\theta})}{1-\tanh^2( \frac{\widetilde{x}}{\theta})} ,
\end{equation}
as
\begin{equation}\label{3.11}
\frac{\rmd^2\widetilde{x}}{\rmd t^{2}}=- \mathcal{V}^{\prime}_{\text{eff}}(
\widetilde{x})- \frac{1}{2\alpha\theta}R(\widetilde{x})\, \frac{\rmd\widetilde{x}}{\rmd t}.
\end{equation}
This approximate evolution equation gives the dynamics of the macroscopic value $\widetilde{x}$ of the oscillator position for fast spins: the nonlinear friction term drives the system towards equilibrium, which corresponds to the  minima of ${\cal V}_{\text{eff}}$. Both the ``renormalization'' of the potential to ${\cal V}_{\text{eff}}$ and the nonlinear friction term are a consequence of the coupling between the oscillator and the internal degrees of freedom. Equation (\ref{3.11}) ceases to hold as $\theta\to 0^+$ because  the spin relaxation time given by (\ref{3.7e}) diverges. A detailed discussion on this point can be found in refs.~\cite{PByC10,BPyC10}. In the remainder of the paper, we will restrict ourselves to a temperature range for which the spins change rapidly compared to the oscillator motion and eq. (\ref{3.11}) holds.

\subsection{\label{s3a}Metastability region}

For $H=0$ and subcritical temperatures, $\theta<1$, the effective potential has two equally deep minima at the symmetric positions $\pm x_{\text{eq}}^{(0)}$ of Section \ref{s2}, and a maximum at $x=0$. For $|H|<H_b$ (see figure \ref{fig4} for a qualitative picture), the effective potential has two minima at $x_R>0$ and $x_L<0$ and a metastability region appears. The globally stable position satisfies $x_iH>0$ ($i=R,L$), while the other one is a metastable state in the thermodynamic sense. It must be stressed that this bistability is present for all the subcritical temperatures $\theta<1$, it is not limited to a region near the critical temperature.

Let us analyze in more detail the situation for weak fields. Expansion of Eq. (\ref{2.4}) in powers of $H=F-F_c\ll 1$ gives
\begin{equation}\label{3.12}
    x_{R,L}=\pm x_{\text{eq}}^{(0)}+\chi H+\mathcal{O}(H^2),
\end{equation}
where $x_{\text{eq}}^{(0)}$ is given by the solution of Eq. (\ref{2.4}) for zero field, and $\chi$ is the zero-field ``susceptibility''
\begin{equation}\label{3.13}
    \chi=\left.\partial_H x_{\text{eq}}\right|_{H=0}=\left(1-\frac{1}{\theta}+
    \frac{{x_{\text{eq}}^{(0)}}^2}{\theta}\right)^{-1}.
\end{equation}
Therefore, $x_R-x_L= 2x_{\text{eq}}^{(0)}$ to lowest order in $H$. Close to the critical temperature, $\theta\to 1^-$, substitution of Eq. (\ref{2.5}) into (\ref{3.13}) yields $\chi\sim\theta/[2(1-\theta)]$. On the other hand, the unstable position $x_U$ changes from zero to
\begin{equation}\label{3.14}
    x_U=-\left( \frac{1}{\theta}-1 \right) H +\mathcal{O}(H^2).
\end{equation}
According to (\ref{2.2}), the transitions between the two minima are hindered by the presence of large energy barriers,
\begin{equation}\label{3.14b}
   B_{R,L}=(N+1)\left[\mathcal{V}_{\text{eff}}(x_U)-\mathcal{V}_{\text{eff}}(x_{R,L})\right],
\end{equation}
which  are proportional to the system size $N+1$. For $H=0$, both equilibrium states have the same energy barrier,
\begin{equation}\label{3.14c}
    B^{(0)}=-(N+1)\mathcal{V}_{\text{eff}}(x_{\text{eq}}^{(0)})>0.
\end{equation}
For $H\neq 0$, the barrier corresponding to the stable equilibrium point verifying  $x_iH> 0$ ($i=R,L$) is larger than the one for the metastable equilibrium point. The barrier from the metastable state tends to zero as $|H|\to H_b$ because $x_U$ and the metastable equilibrium position ($x_L$ for $H>0$, $x_R$ for $H<0$) coalesce in that limit. The maximum effective potential at $x_U$ separates the basins of attraction of the minima at $x_L$ and $x_R$. The system spends long periods of time oscillating in the vicinity of either $x_L$ or $x_R$ until it is able to hop to the other minimum via a thermally activated process. The residence times in each basin of attraction increase exponentially with $N+1$, and we have to consider a system of moderate size in order to see hopping between the two minima on a reasonable time scale.

Let us estimate the barrier height  for weak fields. From Eqs. (\ref{2.3}) and (\ref{3.12}), we have
\begin{equation}\label{3.15}
   \mathcal{V}_{\text{eff}}(x_{R,L})=
   \mathcal{V}_{\text{eff}}(x_{\text{eq}}^{(0)})\mp H x_{\text{eq}}^{(0)} +\mathcal{O}(H^2),
\end{equation}
while $\mathcal{V}_{\text{eff}}(x_U)=\mathcal{O}(H^2)$. In Eq. (\ref{3.15}) and for the rest of this section, the upper sign corresponds to $x_R$ and the lower sign to $x_L$.  The respective barriers from the equilibrium states $x_{R,L}$ defined in Eq. (\ref{3.14b}) are
\begin{equation}
    B_{R,L}\simeq B^{(0)}\pm (N+1)H x_{\text{eq}}^{(0)}.
    \label{3.16}
\end{equation}
The residence times in the respective basins of the minima should have the Arrhenius form,
\begin{equation}\label{3.17}
    \tau_{R,L}=\tau_0 \exp(B_{R,L}/\theta),
\end{equation}
where $\tau_0$ is some characteristic time. By inserting Eq. (\ref{3.16}) into (\ref{3.17}), we get
\begin{equation}\label{3.18}
    \tau_{R,L}=\tau_c \exp\left(\pm \frac{(N+1)x_{\text{eq}}^{(0)}}{\theta} H\right),
\end{equation}
where $\tau_c=\tau_0 \exp(B^{(0)}/\theta)$ is the residence time for zero field in each basin (the same for them both). Equation (\ref{3.18}) is the main result of this section; we should stress that it is valid for weak fields $H\ll 1$, but $(N+1)H$ can be of the order of unity or even a large number. For $H>0$, we have $\tau_R > \tau_L$, because $x_R$ is the globally stable state, while for $H<0$ it is $\tau_R<\tau_L$, since $x_R$ is the metastable state in that case. Interestingly, the ratio of the average lifetimes $\tau_R/\tau_L$ gives the so-called  equilibrium constant $K$ for  folding/unfolding at that force value \cite{LOSTyB01}. Therefore, for our model we arrive at
\begin{equation}\label{3.19}
    K\equiv \frac{\tau_R}{\tau_L}=\exp\left( \frac{2(N+1)x_{\text{eq}}^{(0)}}{\theta} H\right), \quad H=F-F_c,
\end{equation}
$\ln K$ is a linear function of the applied force $F$. A completely analogous behavior has been observed experimentally \cite{LOSTyB01}. The exponent in Eq. (\ref{3.19}) is simply $N+1$ times the difference of values of the effective potential (which plays the role of the free energy per particle) between both states, as readily seen by making use of Eq. (\ref{3.15}).

\section{\label{s4}Numerical results}

We have performed Monte Carlo simulations of the system dynamics introduced above. In all the cases presented here, the dimensionless temperature has been chosen to  be $\theta=0.9$. The qualitative shape of the equilibrium $H$ vs. $x$ curve is similar to the one shown in Fig. \ref{fig4}. For $\theta=0.9$, the oscillator rest position at equilibrium and  zero field is $\pm x_{\text{eq}}^{(0)}\simeq \pm 0.525$, as given by eq. (\ref{2.4}). Interestingly, the approximation in eq. (\ref{2.5}), which is expected to be valid very close to the critical temperature, gives quite a good estimate $x_{\text{eq}}^{(0)}\simeq 0.547$. The points at which the curve $H$ vs. $x$ has either a maximum or a minimum are $\pm x_b=\pm 0.295$; the corresponding values of the field are $\mp H_b=\mp 2.15\times 10^{-2}$. There is metastability for applied fields in the interval $|H|<H_b$.

\subsection{\label{s4a}Force-extension curves}

Let us analyze the force extension curves of the model. The pulling/pushing cycle is as follows. We start at the equilibrium configuration corresponding to $x_{\text{min}}<0$ (folded state), so that the initial value of the field is $H_{\text{min}}=x_{\text{min}}-\tanh(x_{\text{min}}/\theta)$  given by eq. (\ref{2.4}). We pull the system by a stepwise increment of the field: at each step, the field is increased by $\Delta H$, and then the system is allowed to evolve during a given time $\Delta t$. At the end of this period, we record the oscillator position $x$. Then we increase again the field by $\Delta H$ and continue the process in the same vein. The pulling process ends when we reach a positive value of the field $H_{\text{max}}=-H_{\text{min}}$. Then, we start to push back, decreasing the field by $\Delta H$ at each step, until we reach again the minimum field value $H_{\text{min}}$. As during the pulling process, we record the value of $x$ at fixed $H$ after the evolution time $\Delta t$. This process is completely analogous to that carried out in unzipping experiments with biomolecules.

\begin{figure}
\includegraphics[width=\linewidth]{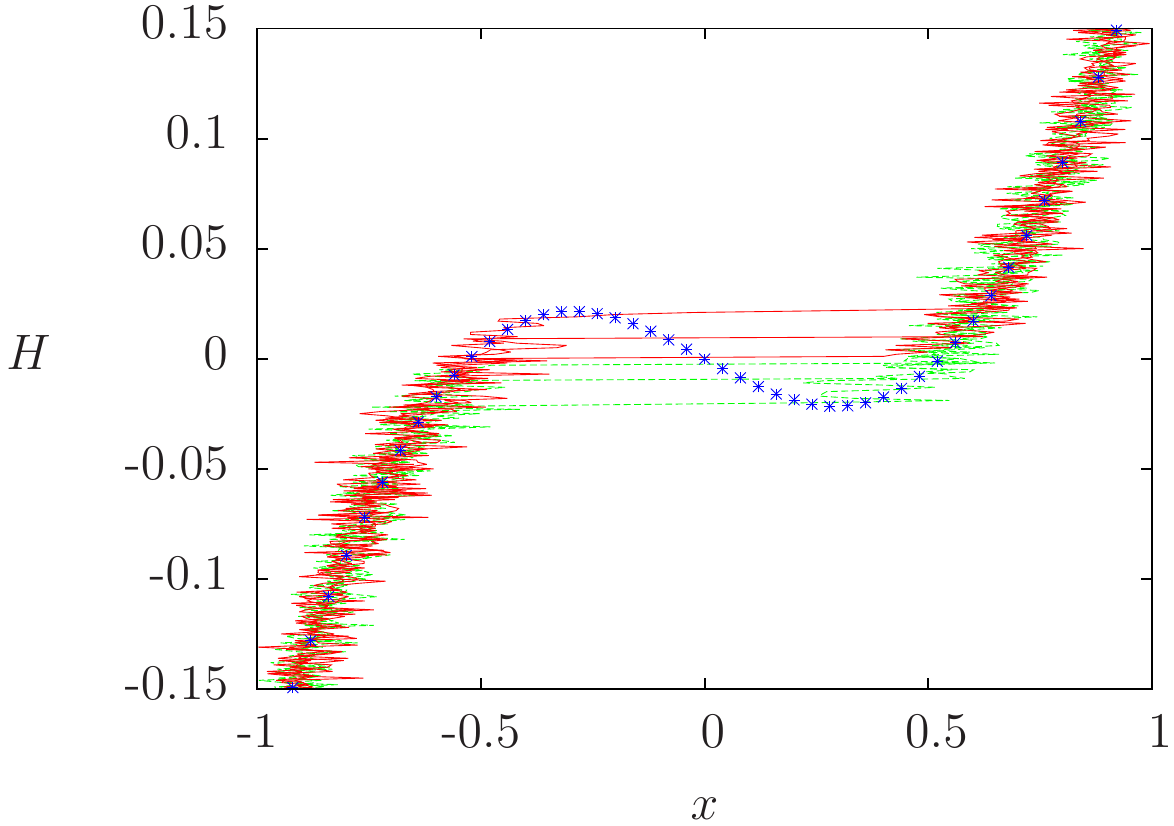}
\caption{\label{fig5} (Color online) Hysteresis cycles for different values of the loading rate $\Delta H/\Delta t$. The red solid lines correspond to the unfolding  process ($\Delta t=5$, $100$ and $10^4$ from top to bottom), and the green dashed lines to the refolding process ($\Delta t=5$, $100$ and $10^4$ from bottom to top). The $H$ versus $x$ curve at equilibrium is plotted with blue stars.}
\end{figure}

Figure \ref{fig5} shows some typical pulling/pushing cycles with different loading rates $\Delta H/\Delta t$. We have used a system with $N+1=1000$ spins, and a spin attempt rate $\alpha=4$, large enough for the spins to be fast as compared to the oscillator \cite{PByC10}. For all the curves, the minimum value of the oscillator position is $x_{\text{min}}=-1.5$, and the field increment at each step is $\Delta H=10^{-3}$, which is smaller than $H_b$ and it allows the system to visit the metastability region. The system behavior is qualitatively similar for other parameter values, as long as the temperature $\theta<1$. The loading rate is changed by varying the amount of time $\Delta t$ at each step. The red (solid) lines correspond to the unfolding process ($\Delta t=5$, $100$ and $10^4$ from top to bottom), and the green (dashed) lines to the refolding process ($\Delta t=5$, $100$ and $10^4$ from bottom to top). These numerical curves are qualitatively similar to those observed in unzipping experiments with nucleic acids \cite{LOSTyB01}. There is always some hysteresis as the unfolding and folding curves are not superimposed. The area of the  hysteresis cycle increases with the loading rate, being large for the largest loading rate considered and almost zero for the smallest one. The main difference with the experimental results is that, in our model, the extension of the molecule is not linked to a dropping of the loading force (in order to see this effect in a real experiment, see for instance figs. 2(A) and 2(E) of ref.~\cite{LOSTyB01}). In the experiments of Ref.~\cite{LOSTyB01}, the total length between the beads localizing the molecule is controlled. This corresponds to fixing the length $L$ in our model, not the load given by $H$ as we have done. When the force is externally controlled in experiments, there is no drop of the loading force at the extension transition and the hysteresis cycle described by the molecule extension is completely analogous to ours \cite{Hu10}.

Our numerical results can be explained as follows. As depicted in Fig. \ref{fig4} and observed in
the last paragraph of Sec. \ref{s2b}, for $|H|>H_b$ we have a unique stable equilibrium point for $|H|>H_b$, the zipped state $x_{\text{eq}}<-x_b<0$ for $H<-H_b<0$ and the unzipped state $x_{\text{eq}}>+x_b>0$ for $H>H_b$. For $|H|<H_b$, there are two (locally) stable rest positions $x_L<0$ and $x_R>0$ and an unstable position $x_U$ between them. The true thermodynamic equilibrium state of the system corresponding to the global minimum of the potential satisfies $x_iH>0$ ($i=L,R$) whereas the metastable state has rest position such that $x_iH<0$.

Let us consider again the pulling processes in fig. \ref{fig5},  starting from the equilibrium configuration corresponding to a low value of the applied force, $H<0$, $|H|\gg |H_b|$. {When we increase the force at a moderate rate, the oscillator follows the equilibrium curve for negative values of $H$, with $x_{\text{eq}}<0$, because its relaxation time is small compared to $\Delta t$, and the friction term in eq. (\ref{3.11}) can drive it to equilibrium. When the field reaches $H=0$ (or a very small value), the stable equilibrium position of the oscillator changes discontinuously from $-x_{\text{eq}}^{(0)}$ to $x_{\text{eq}}^{(0)}$. The important question is now whether $\Delta t$ issufficiently long for the oscillator position to overcome the energy barrier $B^{(0)}$ at zero field, given by Eq. (\ref{3.14c}). If the answer is positive, the system jumps during the time interval $\Delta t$ to the other stable branch where $x>0$, and it stays on that branch when the force is further increased. In the pushing back experiment, the system reverses its path and the behavior is almost reversible. This behavior is observed for the largest value $\Delta t=10^4$, corresponding to a loading rate $\Delta H/\Delta t=10^{-7}$. For larger pulling rates, such as the other ones considered in the same figure, the system does not have enough time to jump over the energy barrier at $H=0$. Therefore, $x$ moves over the metastable branch with $x_iH<0$, until the barrier decreases sufficiently for the oscillator position to jump to the most stable branch, with $x>0$. Of course, the actual part of the metastable region visited by the system depends on the loading rate. For the highest loading rate considered, corresponding to $\Delta t=5$, the system visits the whole metastable branch up to the maximum. A similar line of reasoning explains the behavior observed in the refolding curve. It is interesting to note that the hysteresis cycle found for high loading rates is not a non-equilibrium behavior, as previously suggested \cite{LOSTyB01,LDSTyB02,Ri03,CRJSTyB05}, but it arises from the sampling of the regions of metastability for subcritical temperatures.

A pulling experiment corresponding to a rate  even slower than the smallest loading rate in fig.\ \ref{fig5} is plotted in figure \ref{fig6}. Again, the minimum value of the oscillator position has been chosen to be $x_{\text{min}}=-1.5$, and the field increment at each step $\Delta H=10^{-3}$, but the time spent by the system at each value of the force is very large, namely $\Delta t=10^5$. For the sake of clarity, the region around the critical force $F=F_c$ ($H=0$) has been zoomed in. We observe several jumps between the folded ($x<0$) and unfolded ($x>0$) states for $|H|=|F-F_c|<5\times 10^{-3}$. This behavior is a clear signature of the bistability shown by our system in the region $|H|<H_b$. At zero field, the energy barriers between the two stable states are identical, $\tau_R=\tau_L=\tau_c$ in (\ref{3.18}), and the reverse transition is equally likely. For $\Delta t\gg\tau_c$, the oscillator position has enough time to surpass the energy barriers several times and it may go back and forth from one state to the other, as shown in Fig. \ref{fig6}.
\begin{figure}
\includegraphics[width=\linewidth]{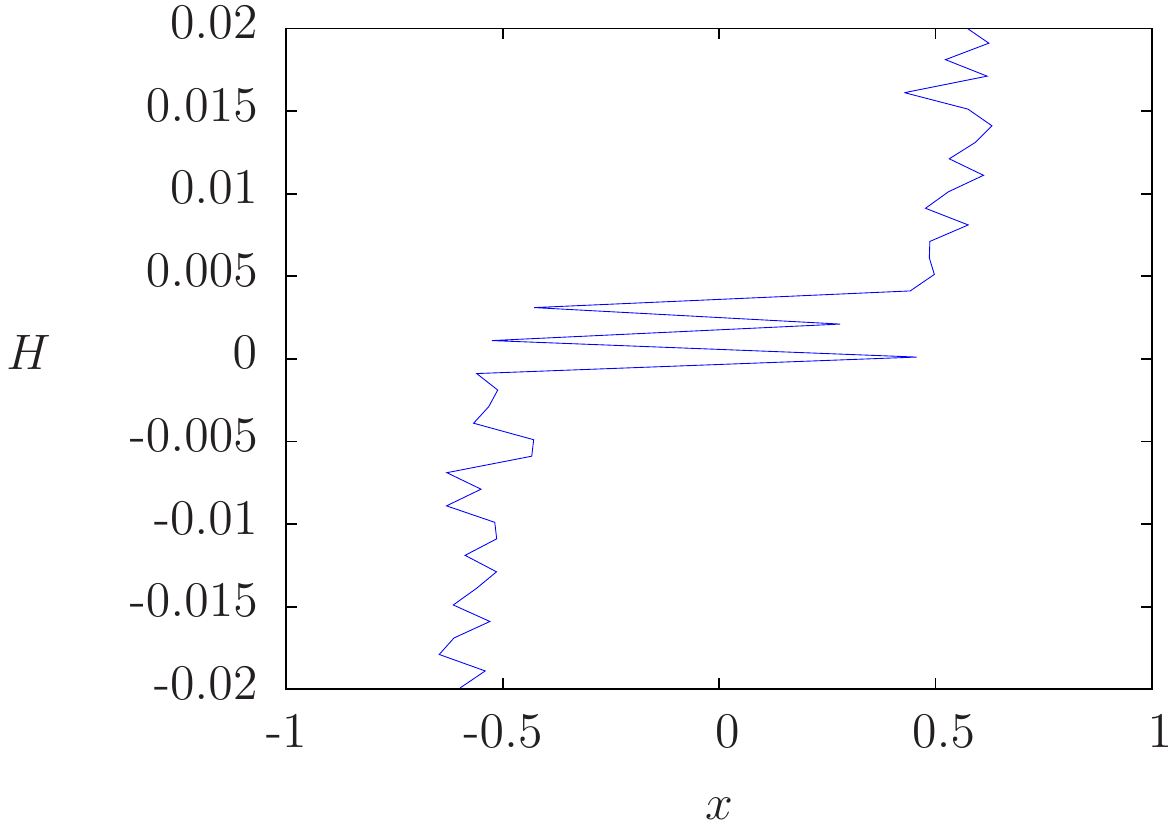}
\caption{\label{fig6} (Color online)  Detail of the metastability  region $|H|\leq H_b$ for a pulling experiment with a very slow loading rate, $\Delta H=10^{-3}$ and $\Delta t=10^5$. Hopping between the zipped and unzipped state is clearly seen for $|H|=|F-F_c|<5\times 10^{-3}$.}
\end{figure}

\subsection{\label{s4b}Constant force experiments}

We have also carried out Monte Carlo simulations at constant force. We have chosen $\theta=0.9$, a small value of the field, $H=3\times 10^{-3}<H_b$, and a smaller size, $N+1=500$, in order to keep the simulation time under control. Equation (\ref{2.4}) gives the two locally stable oscillator rest positions $x_L=-0.509$ (metastable) and $x_R=0.540$ (globally stable), separated by the unstable oscillator position $x_U=-0.027$. These values agree with the weak field expressions (\ref{3.12})-(\ref{3.14}). Figure \ref{fig7} shows a time trace of the system. The oscillator jumps stochastically between the two values $x_L$ and $x_R$ corresponding to the two locally stable equilibrium points.  The system spends more time in the stable state $x_R$ (recall $H=3\times 10^{-3}>0$), because it has to surpass a larger energy barrier in order to escape therefrom.
\begin{figure}
\includegraphics[width=\linewidth]{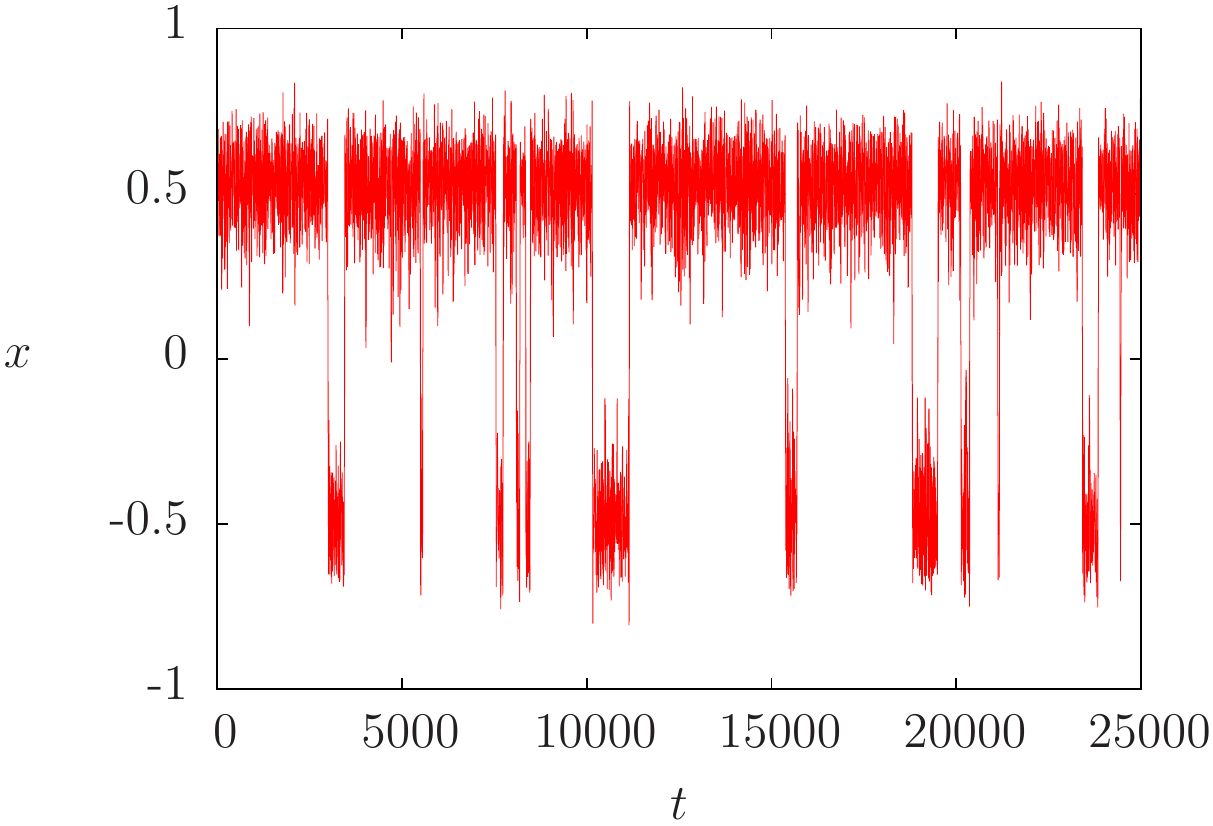}
\caption{\label{fig7} (Color online) Time trace of the oscillator position for a constant force experiment in the metastability region, namely $H=3\times 10^{-3}$. Hopping between the two locally stable equilibrium points of the oscillator is observed.}
\end{figure}

In order to check Eqs. (\ref{3.18}) or (\ref{3.19}) for the residence times in each basin of attraction, we have measured the average time spent in each basin for different values of the applied field in the metastability region $|H|<H_b$. We find that $\ln\tau_{R,L}$ increases linearly with $H$, with a slope that agrees with the theoretical prediction of Eq. (\ref{3.18}). We have plotted the ratio of average lifetimes $K=\tau_R/\tau_L$ defined in Eq. (\ref{3.19}) as a function of the applied field $H$ in Fig. \ref{fig8}. Therein, $\ln K$ shows a linear behavior, similarly to that seen in actual experiments \cite{LDSTyB02}. The slope $m=\rmd \ln K/\rmd H$ obtained numerically, $m=542.5$ agrees well with the theoretical prediction calculated from Eq. (\ref{3.19}), $m=583.8$. This result strongly supports the physical picture developed in Section \ref{s3a} for the behavior of the system in the metastability region, both from a qualitative and a quantitative point of view.

\begin{figure}
\includegraphics[width=0.95\linewidth]{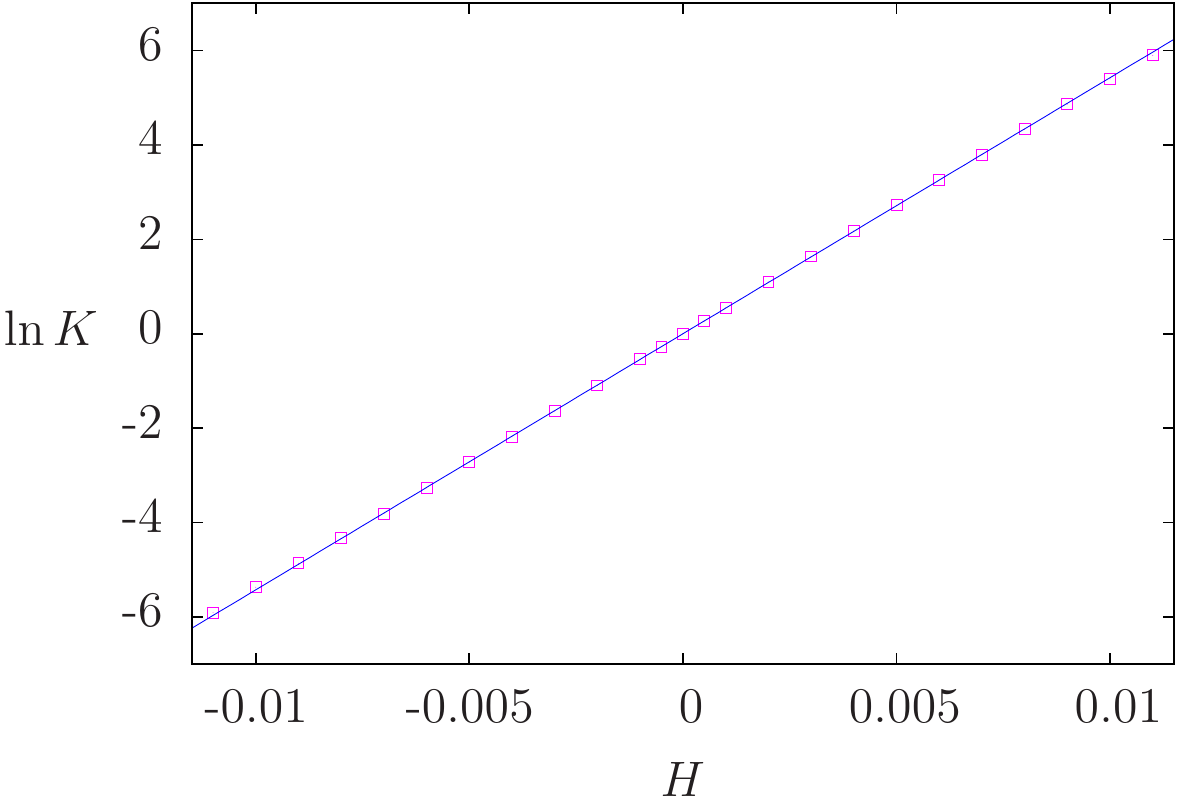}
\caption{\label{fig8} (Color online) Logarithm of the ratio $K\equiv\tau_R/\tau_L$ (crosses) of the residence times for the two equilibrium states $x_{R,L}$ in the metastability region. The number of spins is $N+1=500$. Also plotted is the best fit to the Arrhenius law (solid line), Eq. (\ref{3.19}). }
\end{figure}

\section{\label{s5}Conclusions}

We have modeled DNA folding/unfolding under an external load force by a macroscopic linear oscillator coupled with internal degrees of freedom represented by Ising spins that undergo Glauber dynamics. The simple mean-field character of the model  prevents us from doing quantitative comparisons with the real experiments, and we have to settle for qualitative comparisons. We cannot simulate position-controlled experiments, only force-controlled ones.

Despite its limitations, the picture arising from this simple model is physically appealing.
The hysteresis cycles show that the system exhibits a metastable equilibrium behavior in the unzipping experiments, not a true non-equilibrium behavior, as it was suggested previously \cite{LOSTyB01,LDSTyB02,Ri03,CRJSTyB05}. In this regard, the unfolding/refolding cycles are quite different from the truly nonequilibrium hysteresis cycles exhibited by glass formers in cooling/heating processes (see \cite{BPyR94} and references therein). In the cooling process, the glass formers depart from the equilibrium curve and end in a far from equilibrium state at low temperatures. In the reheating process, they return to equilibrium following a curve different from the cooling one, which typically overshoots the equilibrium curve. In contrast with this behavior, the hysteresis cycles in our system arise because the metastable equilibrium branches of the $x$ vs $H$ curve are swept at high loading rate (such that the system does not have enough time to find the true minimum of the potential in the bistable field interval $|H|<H_b$). Due to the shape of the equilibrium $x$ vs $H$ curve in fig. \ref{fig4}, the system unzips at a higher value of the field than the one at which it rezips. We expect that this general picture remains valid for more realistic models and/or actual biomolecules.

As in the experiments, the system hops between the zipped and the unzipped state at very small loading rates. It has then enough time to surpass the energy barrier separating the unfolded and the folded states. When the force is held constant with $|H|<H_b$ (bistable region), the system hops back and forth between the two possible equilibrium values of the oscillator position. The average lifetimes show an Arrhenius-like dependence on the applied field $H=F-F_c$, again in agreement with the behavior of real systems.

For all subcritical temperatures $T<T_c$, these behaviors occur due to the first order transition
and its associated region of metastability. The phase transition is a consequence of the coupling between the oscillator and the 1d spin system, which introduces an effectively long-range interaction between the spins. Similar hidden 1d long range effective correlations enable phase transitions in biological systems \cite{Pe06}.

\acknowledgments

This research has been supported by the Spanish Ministerio de Ciencia e
Innovaci\'on (MICINN) through Grants FIS2011-28838-C02-01 (LLB),
FIS2011-28838-C02-02 (AC), FIS2011-24460 (AP, partially financed by FEDER
funds) and FIS2010-22438-E (Spanish National Network Physics of
Out-of-Equilibrium Systems) and by UCM/BSCH CM 910143 (AC).

\end{document}